\documentclass[twocolumn,showpacs,preprintnumbers]{revtex4}
\usepackage{amsmath}
\usepackage[dvips]{graphicx}
\usepackage{bm}

\begin{document}

\title{Vortex core size in interacting cylindrical nanodot arrays}
\author{D. Altbir$^1$}
\author{J. Escrig$^1$}
\author{P. Landeros$^2$}
\author{F. S. Amaral$^3$}
\author{M. Bahiana$^3$}
\affiliation{$^{1}$Departamento de F\'{\i}sica, Universidad de Santiago de Chile, USACH,
Av. Ecuador 3493, Santiago, Chile\\
$^{2}$Departamento de F\'{\i}sica, Universidad T\'{e}cnica Federico Santa Mar%
\'{\i}a, Av. Espa\~{n}a 1680, Valpara\'{\i}so, Chile\\
$^{3}$Instituto de F\'{\i}sica, Universidade Federal do Rio de
Janeiro, Caixa Postal 68.528, 21941-972, RJ, Brazil}
\date{\today}

\begin{abstract}
The effect of dipolar interactions among cylindrical nanodots,
with a vortex-core magnetic configuration, is analyzed by means of
analytical calculations. The cylinders are placed in a $N\times N$
square array in two configurations - cores oriented parallel to
each other and with antiparallel alignment between nearest
neighbors. Results comprise the variation in the core radius with
the number of interacting dots, the distance between them and dot
height. The dipolar interdot coupling leads to a decrease
(increase) of the core radius for parallel (antiparallel) arrays.
\end{abstract}

\maketitle

\section{Introduction}

Regular arrays of magnetic particles produced by nanoimprint lithography
have attracted strong attention during the last decade. Common structures
are arrays of wires, \cite{V01, NHW+02} cylinders, \cite{CKA+99, RHS+02}
rings, \cite{CRE+04, RKL+01} and tubes. \cite{NCR+05, NCM+05} Such
structures can be tailored to display different stable magnetized states,
depending on their geometric details. Besides the basic scientific interest
in the magnetic properties of these systems, there is evidence that they
might be used in the production of new magnetic devices or as media for high
density magnetic recording. \cite{Chou97} In particular, two-dimensional
arrays of magnetic nanoparticles have been proposed as candidates for
magnetoresistive random access memory (MRAM) devices. \cite{Daughton99,
ZZP00, Parkin04}

Recent studies on such structures have been carried out with the
aim of determining the stable magnetized state as a function of
the geometry of the particles. \cite{CKA+99, RHS+02, AAR+02,
LEA+05, PH04, ELA+06} In the case of cylindrical particles, three
idealized characteristic configurations have been identified:
ferromagnetic with the magnetization parallel to the basis of the
cylinder, ferromagnetic with the magnetization parallel to the
cylinder axis, and a vortex state in which most of the magnetic
moments lie parallel to the basis of the cylinder. The occurrence
of each of these configurations depends on geometrical factors,
such as the linear dimensions and their aspect ratio $\tau \equiv
H/R$, with $H$ the height and $R$ the radius of the cylinder.
\cite{CKA+99, RHS+02, AAR+02, LEA+05} Another issue to be
considered is the interaction between particles. In this case the
interparticle distance, $D$, is the important parameter.
\cite{Metlov06, LEL+07,EAJ+07} Usually $D$ is large enough to make
the exchange coupling between particles negligible, making the
dipolar interaction a fundamental point concerning the magnetic
state of the system. \cite{ELA+06, PH05}

In this paper we focus on cylindrical particles with dimensions
such that, in the absence of an external field, the magnetic state
is a vortex. For this magnetic configuration the dipolar
interaction between the dots is due to the existence of a core
region \cite{SOH+00, WWB+02} in which the magnetic moments have a
nonzero component parallel to the cylinder axis. In this case it
is relevant to understand how the core magnetization is affected
by the interparticle interaction. Recently, Porrati \textit{et
al.} \cite{PH05} investigated an array of dots using micromagnetic
simulations. In small arrays they observed that the dipolar
interaction changes the size of the magnetic core. However, and
because of the use of micromagnetic simulations, larger arrays
have not been investigated. Therefore, analytical calculations are
very desirable to compare with experiments. With this in mind we
examine the behavior of the core in arrays of dots in the
vortex-core magnetic phase, in configurations with parallel and
antiparallel alignment between the cores. We consider analytical
calculations based on a continuous description of the
dots.\cite{Aharoni96}

\section{System \& Units}

The basic parameters and variables used in our calculations are
summarized in figure 1. We consider cylindrical dots with radius
$R$ and height, or thickness, $H$, in square arrays with $N\times
N$ dots and center-to-center lattice spacing $D$. The distance
between any two dots in the array is
denoted by $S$. Whenever necessary, cylindrical coordinates $\rho $ and $%
\phi $ are defined on the dot plane normal to the cylindrical $z$-axis.

The experimental measurement of the vortex core profile and core radius is
not a simple task, and usually it is the full core magnetization, $\mu _{z}$%
, which is measured. \cite{SOH+00,RSpc} With this in mind, the
vortex core
is characterized through the calculation of an effective core radius, $C_{%
\mbox{\it\scriptsize eff}}$, defined as the radius of an effective cylinder,
uniformly magnetized along its axis, and whose total magnetic moment,
\begin{equation}
\mu _{z}=M_{0}V=M_{0}\pi HC_{\mbox{\it\scriptsize eff}}^{2},  \label{muz}
\end{equation}%
is the same as produced by the $z$-component of the vortex core,
as depicted in figure 1. Here $V$ is the dot volume and $M_{0}$ is
the saturation magnetization.

Two types of magnetic ordering within the array are examined: all
cores parallel (configuration P) and anti-parallel
nearest-neighbor cores (configuration AP). This choice is not an
arbitrary one, because the P configuration corresponds to the
saturated one and the AP configuration is the ground state of the
array.

\begin{figure}[h]
\begin{center}
\includegraphics[width=8cm,height=4cm]{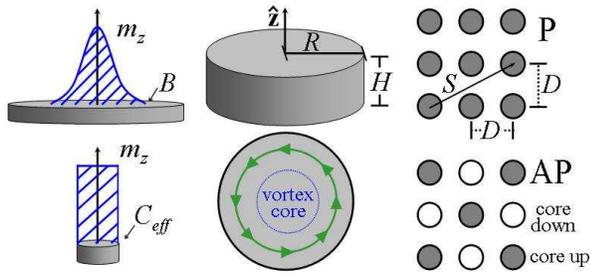}
\end{center}
\caption{An illustration of geometrical parameters defining each
dot and the array.}
\end{figure}

All linear dimensions will be considered in units of the exchange length $L_{%
\mbox{\it\scriptsize x}}$, defined as $L_{\mbox{\it\scriptsize x}}=\sqrt{%
2A/\mu _{0}M_{0}^{2}}$. The dimensionless geometrical parameters are then
defined as
\begin{equation}
h\equiv \frac{H}{L_{\mbox{\it\scriptsize x}}},\ r\equiv \frac{R}{L_{%
\mbox{\it\scriptsize x}}},\ d\equiv \frac{D}{L_{\mbox{\it\scriptsize
x}}},\ b=\frac{B}{L_{\mbox{\it\scriptsize
x}}},\ c_{\mbox{\it\scriptsize eff}}\equiv \frac{C_{%
\mbox{\it\scriptsize
eff}}}{L_{\mbox{\it\scriptsize x}}},\ s\equiv \frac{S}{L_{%
\mbox{\it\scriptsize
x}}}\,.  \label{gparameters}
\end{equation}

\section{Theoretical model}

Large arrays can be studied if we adopt a simplified description of the
system in which the discrete distribution of magnetic moments is replaced
with a continuous one, defined by a function $\vec{M}(\vec{r})$ such that $%
\vec{M}(\vec{r})\delta V$ gives the total magnetic moment within the element
of volume $\delta V$ centered at $\vec{r}$. Using such a description, the
internal energy of the array can be written in terms of the self-energies, $%
E_{\mbox{\it\scriptsize self}}$, that is, the energies of the isolated dots,
and the interaction contribution, $E_{\mbox{\it\scriptsize
int}}$, corresponding to the dipolar coupling between the dots. Both $E_{%
\mbox{\it\scriptsize self}}$ and $E_{\mbox{\it\scriptsize int}}$ have a
magnetostatic term given by $E_{\mbox{\it\scriptsize dip}}=(\mu _{0}/2)\int
\vec{M}\cdot \nabla U\,dV$ , with $U(\vec{r})$ the magnetostatic potential.
Assuming that $\vec{M}(\vec{r})$ varies slowly on the scale of the lattice
parameter, the exchange term can be approximated by $E_{%
\mbox{\it\scriptsize
ex}}=A\int \sum (\nabla m_{i})^{2}dV$, where $m_{i}$ is the $i$-th component
of the reduced magnetization with respect the saturation value $M_{0}$, that is, $%
m_{i}=M_{i}/M_{0}$, for $i=x,\ y,\ z$.\cite{Aharoni96}

\subsection{Vortex-core magnetization}

For the vortex-core configuration we assume that the magnetization is
independent of $z$ and $\phi $, that is,
\begin{equation}
\vec{m}(\vec{r})=m_{z}(\rho ){\hat{z}}+m_{\phi }(\rho ){\hat{\phi}}\ ,
\label{mv}
\end{equation}%
\noindent where $\rho $ is the radial coordinate, $\ {\hat{z}}$ and ${\hat{%
\phi}}$ are unitary vectors in cylindrical coordinates, and the
normalization condition requires that $m_{z}^{2}+m_{\phi }^{2}=1$.
The function $m_{z}(\rho )$ specifies the core profile, for which
we adopt the model proposed by Landeros \textit{et
al},\cite{LEA+05} given by
\begin{equation}
m_{z}(\rho )=\left[ 1-(\rho /B)^{2}\right] ^{n}\;,\;\ \ 0<\rho <B\;.
\label{model}
\end{equation}%
\noindent and $m_{z}(\rho )=0$\ if $B<\rho <R$. Here $B$ is a
parameter related to the core radius and the exponent $n$ is a
non-negative integer. Alternative expressions for $m_{z}(\rho )$
have been proposed in the literature. \cite{FT65, UP93, Aharoni90,
HKK03} Figure 2 illustrates the calculated magnetization profile
$m_{z}(\rho )$ for a Fe dot ($R=28.2$ nm and $H=37.6$ nm) using
different models. The (gray) dash-dotted line corresponds to the
model by Usov \textit{et al}\cite{UP93}, the (blue) thin line
corresponds to the one proposed by Aharoni,\cite{Aharoni90} the
(black) thick line corresponds to our model\cite{LEA+05} with
$n=4$, the dashed (red) line represents $m_{z}(\rho )$ using the
model proposed by Feldtkeller \textit{et al}\cite{FT65}, and the
dotted (green) line
represents the profile using the model presented by H\"{o}llinger \textit{%
et al}\cite{HKK03}.

\begin{figure}[h]
\begin{center}
\includegraphics[width=8cm,height=8cm]{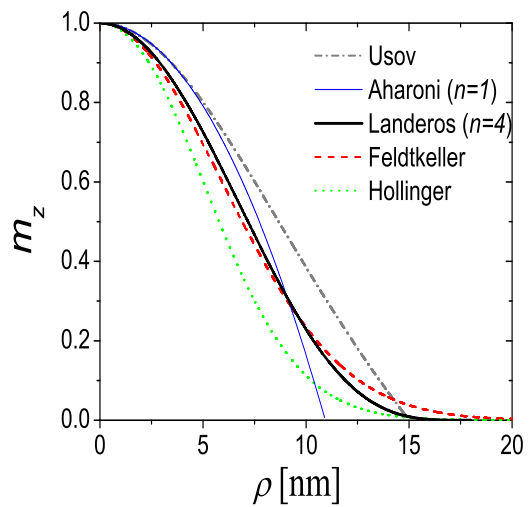}
\end{center}
\caption{Calculated magnetization profiles, $m_{z}(\protect\rho
)$, for different core models. Dimensions are $R=28.2$ nm,
$H=37.6$ nm and the material (iron) es defined by $L_{x}=3.327$
nm.}
\end{figure}

Note that the model proposed by A. Aharoni\cite{Aharoni90} is equivalent to
our model (Eq. \ref{model}) using $n=1$. We can observe that our model with $%
n=4$ agrees well with the ones presented in [25] and [28], and has
no discontinuities at $m_{z}(\rho =B)=0$, as in the models
proposed by Usov \textit{et al}\cite{UP93} and
Aharoni,\cite{Aharoni90} making the vortex core profile easily
integrable. The form for $m_{z}(\rho )$ presented above (Equation
(\ref{model})) allows us to obtain the magnetization profile by
minimizing the total energy of the array. The value of $B$ is
determined in
the minimization process, as explained below, and the effective core radius (%
$C_{\mbox{\it\scriptsize eff}}$) can be evaluated equating the
$z$-component of total magnetic moment (Equation (\ref{muz})) with
the core magnetic moment,
given by%
\begin{equation}
\mu _{z}=M_{0}\int_{V}m_{z}(\rho )\rho \,d\rho \,dz\,d\phi \,.
\end{equation}%
Using the proposed model for $m_{z}(\rho )$, Equation (\ref{model}), we obtain $%
\mu _{z}=M_{0}\pi HB^{2}/\left( n+1\right) $\ and therefore,
\begin{equation}
C_{\mbox{\it\scriptsize eff}}=\frac{B}{\sqrt{n+1}}\;.
\end{equation}

\subsection{Total energy calculation}

We write all the energies in the dimensionless form, $\tilde{E}=E/\mu
_{0}M_{0}^{2}L_{\mbox{\it\scriptsize x}}^{3}$. The energies in the P and AP
configurations will be denoted as $\tilde{E}^{+}$ and $\tilde{E}^{-}$,
respectively, and can be written as%
\begin{equation}
\tilde{E}^{\pm }=N^{2}\tilde{E}_{\mbox{\it\scriptsize self}}+\tilde{E}_{%
\mbox{\it\scriptsize int}}^{\pm }\;,  \label{em}
\end{equation}%
where the first term includes the self-energies of the $N^{2}$
isolated dots, and the second term is the interdot magnetostatic
coupling.

\subsubsection{Self-energy}

We consider two contributions to the self-energy, $\tilde{E}_{%
\mbox{\it\scriptsize self}}$, of an isolate dot in the vortex-core
configuration: the dipolar, $\tilde{E}_{\mbox{\it\scriptsize
dip}}$, and the exchange, $\tilde{E}_{\mbox{\it\scriptsize ex}}$,
contributions. Analytical expressions for the dipolar and exchange
energies have been previously calculated by Landeros \textit{et
al.}\cite{LEA+05} The dipolar
energy results as%
\begin{equation}
\tilde{E}_{\mbox{\it\scriptsize dip}}=\pi \alpha _{n}b^{3}-\frac{\pi \beta
_{n}b^{4}}{4h}F_{n}\left( b/h\right) ,
\end{equation}%
\noindent where
\begin{equation*}
\alpha _{n}\equiv \frac{2^{2n-1}\Gamma (n+1)^{3}}{\Gamma (n+3/2)\Gamma
(2n+5/2)}\,,\quad \beta _{n}\equiv \frac{1}{(n+1)^{2}}
\end{equation*}%
\noindent and%
\begin{equation*}
F_{n}(x)=_{P}F_{Q}[(1/2,1,n+3/2),(n+2,2n+3),-4x^{2}]\,.
\end{equation*}%
\noindent Here, $_{P}F_{Q}$ denotes the generalized hypergeometric
function, $\Gamma \left( x\right) $ is the gamma function and $n$
corresponds to the exponent used in the core model defined by
Equation (\ref{model}).

For integer values of $n$, the exchange energy reduces to \cite{LEA+05}%
\begin{equation}
\tilde{E}_{\mbox{\it\scriptsize ex}}=\pi h\left[ \ln (r/b)+\gamma _{n}\right]
\ ,  \label{eexvm}
\end{equation}%
\noindent where $\gamma _{n}=(1/2)\mathcal{H}\left( 2n\right) -n\mathcal{H}%
\left( -1/2n\right) $. Here, $\mathcal{H}\left( z\right) =\sum_{i=1}^{\infty
}\left[ 1/i-1/(i+z)\right] $ is the generalized harmonic number function\cite%
{EM04} of the complex variable $z$.

\subsubsection{Interdot magnetostatic coupling}

The dipolar interaction between any two dots in the array depends
on the center-to-center distance between them, $S$, and on the
relative orientation of the magnetic cores. An expression for this
energy is obtained using the magnetostatic field experienced by
one of the dots due to the other. Details of these calculations
are included in appendix. The resulting expression is
\begin{equation}
\tilde{E}^{\pm }(S)=\pm \frac{2\pi }{L_{\mbox{\it\scriptsize x}}^{3}}%
\int\limits_{0}^{\infty }dk\,(1-e^{-kH})J_{0}(kS)\left[ \int%
\limits_{0}^{R}J_{0}(k\rho )m_{z}(\rho )\rho d\rho \right] ^{2}\,.
\label{EintMz}
\end{equation}%
\noindent Since the size of the core will be obtained by energy
minimization, in the previous equation we have considered that the
two interacting dots exhibit the same magnetic profile, that is,
their cores are identical. Equation (\ref{EintMz}) allows us to
write the interaction energy between two dots as $\tilde{E}^{\pm
}(S)=\pm \tilde{E}(S)$, depending on the relative orientation of
the cores. Note that if the core size increases,
i.e. $m_{z}$ grows, then $\tilde{E}(S)$ increases. Also we always have $%
\tilde{E}(S)>0$, and the interaction energy between two dots is
always greater than zero ($\tilde{E}^{+}(S)>0$) if the cores are
oriented in the same direction, which causes a shrinking of the
vortex core, in agreement with micromagnetic
simulations\cite{PH05}. On the other hand, if the two cores
are oriented in opposite directions, then the interaction term is negative ($%
\tilde{E}^{-}(S)<0$) and the vortex core expands to lower the
total energy.

Substituting our expression for $m_{z}\left( \rho \right) $, Equation (\ref{model}%
), we obtain
\begin{multline}
\tilde{E}(s)=2^{2n+1}\pi \Gamma ^{2}(n+1)\frac{h^{2n+1}}{b^{2n-2}}
\label{Es1} \\
\int\limits_{0}^{\infty }\frac{dy}{y^{2n+2}}(1-e^{-y})J_{0}\left( y\frac{s}{h%
}\right) J_{n+1}^{2}\left( y\frac{b}{h}\right) \,.
\end{multline}%
\noindent Using equation (\ref{Es1}) and adding up contributions
over the entire
array we obtain the expression for the total interaction energy of the $%
N\times N$ square array as
\begin{multline}
\tilde{E}_{\mbox{\it\scriptsize int}}^{\pm
}(N)=2N\sum_{p=1}^{N-1}(N-p)(\pm 1)^{p}\tilde{E}(pd)  \label{eint9} \\
+2\sum_{p=1}^{N-1}\sum_{q=1}^{N-1}(N-p)(N-q)(\pm 1)^{p-q}\tilde{E}\left( d%
\sqrt{p^{2}+q^{2}}\right) .
\end{multline}%
\noindent This equation was previously obtained by Laroze \textit{et al},
\cite{LEL+07} and has been used to investigate the magnetostatic coupling in
arrays of magnetic nanowires.

At this point we need to specify the value of the parameter $n$ for the core
profile defined by equation (\ref{model}). In a previous work, Landeros \textit{%
et al} \cite{LEA+05} showed that the magnetic vortex core can be
well described for almost any value of $n>1$. We choose $n=4$, as
explained in [14], and finally obtain the following expression for the self-energy,%
\begin{equation}
\tilde{E}_{\mbox{\it\scriptsize self}}=0.0298\pi b^{3}-\frac{\pi b^{4}}{100h}%
F_{4}({b}/{h})+\pi h\left( \ln \frac{r}{b}+2.266\right) ,  \label{self2}
\end{equation}%
\noindent with%
\begin{eqnarray*}
F_{4}(x) &=&-\frac{1}{63x^{10}}(256+384x^{2}+576x^{4} \\
&&+600x^{6}+350x^{8}-256F_{21}(-9/2,1/2,6,-4x^{2}))\,,
\end{eqnarray*}%
where $F_{21}(a,b,c,z)$ is a hypergeometric function.\

Using $n=4$ in Eq. (\ref{Es1}) we obtain%
\begin{equation}
\tilde{E}(s)=294912\frac{\pi h^{9}}{b^{6}}\int\limits_{0}^{\infty }\frac{dy}{%
y^{10}}(1-e^{-y})J_{0}\left( y\frac{s}{h}\right) J_{5}^{2}\left( y\frac{b}{h}%
\right) .  \label{Es2}
\end{equation}

Finally, the total energy of the array is calculated as $\tilde{E}^{\pm
}=N^{2}\tilde{E}_{\mbox{\it\scriptsize self}}+\tilde{E}_{\mbox{\it%
\scriptsize int}}^{\pm }$ , with $\tilde{E}_{\mbox{\it\scriptsize
self}}$\ given by equation (\ref{self2}) and $\tilde{E}_{\mbox{\it\scriptsize int}%
}^{\pm }$\ given by equations (\ref{eint9}) and (\ref{Es2}).

To determine the vortex core magnetization, we have to minimize $\tilde{E}%
^{\pm }$ with respect to $b$. Note that the only term in the expression for
the energy which depends on the radius $r$ of the dot is $\tilde{E}_{%
\mbox{\it\scriptsize self}}$ (see equation (\ref{self2})).
However, the derivative of $\tilde{E}_{\mbox{\it\scriptsize
self}}$ with respect to $b$ is independent of $r$, leading to a
core size that is independent of the dot radius. \cite{LEA+05,
SOH+00} This follows from the fact that the external region of the
dot (a perfect vortex) does not interact with the core (apart from
the exchange interaction across the interface between the two
regions). That is to say, the equation for $b$ that minimizes the
total energy of the vortex configuration is independent of $r$.

\section{Results and discussion}

\begin{figure}[h]
\begin{center}
\includegraphics[width=8cm,height=14cm]{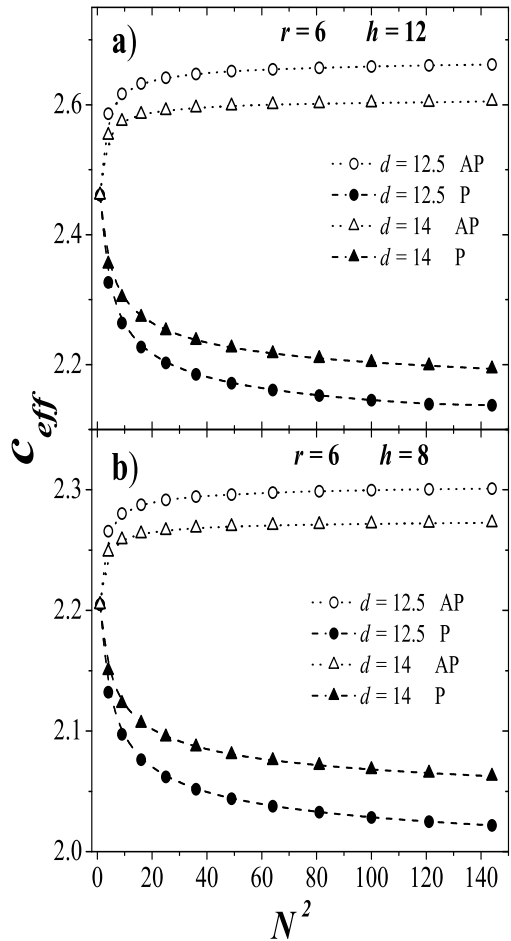}
\end{center}
\caption{Effective core radius $c_{\mbox{\it\scriptsize eff}}$ vs
$N^{2}$ for $r=6$: (a) $h=12$ and (b) $h=8$. Open symbols
correspond to an antiparallel (AP) ordering and full symbols to a
parallel (P) ordering. Circles correspond to $d=12.5$ and
triangles to $d=14$.}
\end{figure}

We are now in position to investigate how the core radius is
affected by the interdot magnetostatic coupling. In order to
obtain the value of $b$, the energy $\tilde{E}^{\pm }$ must be
minimized for fixed $h$, $d$ and $N$. Since the dipolar
interaction is long-ranged, an increment in $N$ leads to an
increase in the dipolar field felt by each dot and, therefore, to
a change in the core radius until a certain asymptotic value is
reached. Figure 3 illustrates this effect for arrays of dots with
$r=6$, $d=12.5$ and $14$, and $h=12$ (figure 3(a)) and $h=8$
(figure 3(b)), in configurations P and AP. For both values of $h$
we observe that for $N=12$ the core radius is almost at its
limiting size. From our results we observe an increase in the core
radius with the number of dots in the AP configuration, and the
opposite behavior for the P configuration, in agreement with
micromagnetic simulations.\cite{PH05} For a given value of $N$, AP
arrays with smaller values of $d$ present larger core radius due
to the preference for the AP ordering. The antiparallel alignment
has the lowest dipolar energy, so the core radius increases with
the number of dots in this configuration. On the other hand, in
the P configuration the parallel coupling between the cores
increases the interaction energy, and then the core region is
reduced in order to decrease the interaction.

The dipolar energy may also be varied if the interdot distance is
changed. This effect is depicted in figure 4 for the AP and P
configurations. While in the antiparallel arrays the size of the
core rapidly reaches the value for isolated dots; in the parallel
configuration, the effect of the interdot interaction is relevant
for longer interdot distances.

\begin{figure}[h]
\begin{center}
\includegraphics[width=8cm,height=14cm]{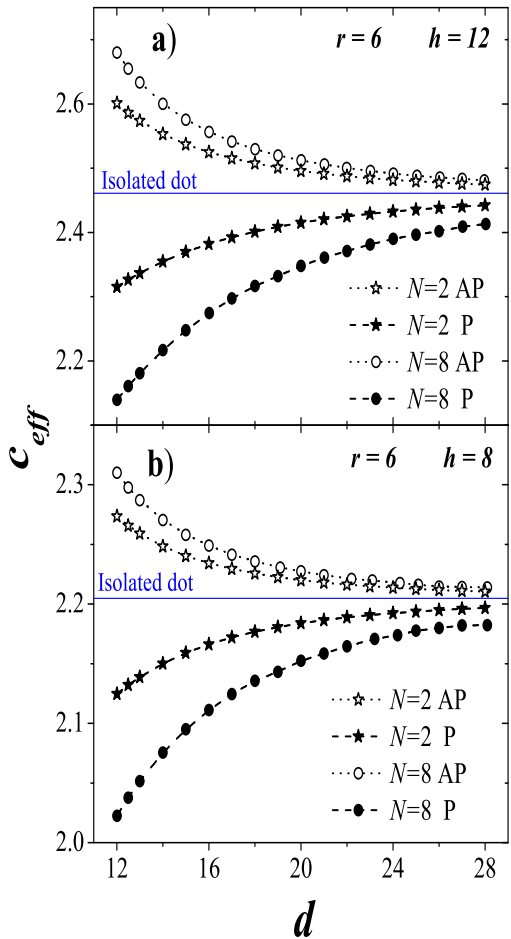}
\end{center}
\caption{Effective core radius $c_{\mbox{\it\scriptsize eff}}$ vs $d$ for $%
r=6$: (a) $h=12$ and (b) $h=8$. Open symbols correspond to an
antiparallel (AP) ordering and full symbols to a parallel (P)
ordering.}
\end{figure}

Now, the influence of the dot height is more subtle. Figure 5
shows the behavior of $c_{\mbox{\it\scriptsize
eff}}$ vs $h$ for arrays of dots with $r=6$, $d=12$, with $N^{2}=4$ and $64$%
, in the P and AP configurations. For the isolated dot (blue solid line),
the effective core follows approximately the relation $c_{%
\mbox{\it\scriptsize eff}}\approx 1.228+0.291h^{0.577}$.

\begin{figure}[h]
\begin{center}
\includegraphics[width=8cm,height=8cm]{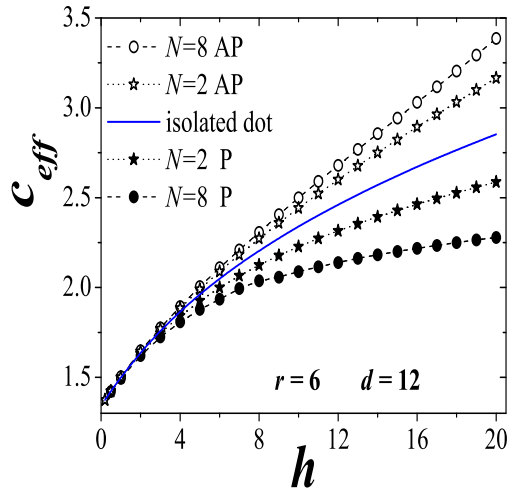}
\end{center}
\caption{Effective core radius $c_{\mbox{\it\scriptsize eff}}$\ vs $h$ for
dots with $d=12$, $r=6$. Open symbols correspond to an antiparallel (AP)
ordering and full symbols to a parallel (P) ordering. Stars correspond to $%
N^{2}=4$ and circles to $N^{2}=64$.}
\end{figure}

For an isolated dot, a transition from the vortex configuration to
a complete ferromagnetic ordering along the dot axis is observed
as $h$ increases, as shown in the phase diagrams presented in
[14]. From the point of view of the core radius, this transition
may be seen as a slow and
continuous increase in $c_{\mbox{\it\scriptsize eff}}$ with $h$, until $c_{%
\mbox{\it\scriptsize eff}}\approx r$. As the dots interact in the
array this behavior may change. To illustrate this point we
calculate the transition line from the out-of-plane uniform state
to the vortex-core state configuration in arrays of 4 and 64 dots.
The self-energy for the out-of-plane uniform ($u$) state has been
presented in \cite{TBZ+04}
and reads%
\begin{equation}
\tilde{E}_{\mbox{\it\scriptsize self}}^{u}=\frac{\pi hr^{2}}{2}\left( 1+%
\frac{8r}{3\pi h}-F_{21}\left[ -\frac{4r^{2}}{h^{2}}\right] \right) \ ,
\label{EselfU}
\end{equation}%
\noindent where $F_{21}[x]=F_{21}[-1/2,1/2,2,x]$ is a hypergeometric
function.

The interaction energy between two dots with full magnetization
along their axis can be obtained from equation (14) using
$m_{z}(\rho )=1$ and gives
\begin{equation}
\tilde{E}^{u}(s)=2\pi hr^{2}\int\limits_{0}^{\infty }\frac{dy}{y^{2}}%
(1-e^{-y})J_{0}(y\frac{s}{h})J_{1}^{2}(y\frac{r}{h})\,,  \label{Eintunif}
\end{equation}%
as presented by Beleggia \textit{et al.}\cite{BTZ+04} The total energy of an
array with out-of-plane uniform magnetization is given by%
\begin{equation}
\tilde{E}^{u\pm }=N^{2}\tilde{E}_{\mbox{\it\scriptsize self}}^{u}+\tilde{E}_{%
\mbox{\it\scriptsize int}}^{u\pm }\;,  \label{Eu}
\end{equation}%
with the above expressions for the self-energy (equation
(\ref{EselfU})) and the interaction energy
($\tilde{E}_{\mbox{\it\scriptsize int}}^{u\pm }$) given by
equation (\ref{eint9}) with the magnetostatic coupling of the two
dots given by equation (\ref{Eintunif}). The transition lines
between the vortex-core and uniform out-of-plane magnetic states
are obtained by equating equations (\ref{em}) and (\ref{Eu}) and
are represented in figure 6.

\begin{figure}[h]
\begin{center}
\includegraphics[width=8cm,height=8cm]{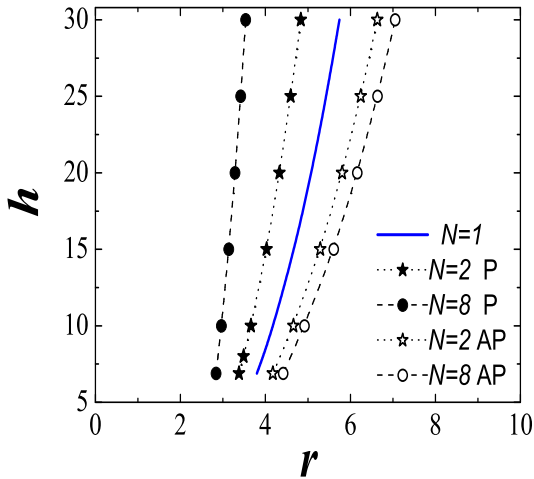}
\end{center}
\caption{Dipolar induced shifts in the transition line that separates the
vortex and the out-of-plane uniform magnetic states. We have fixed the ratio
$r/d=0.48$. Open symbols correspond to an antiparallel (AP) ordering and
full symbols to a parallel (P) ordering. Stars correspond to $N=2$ and
circles to $N=8$.}
\end{figure}

As we expected, the transition line shifts to lower radii (and greater
heights) in the parallel array and to bigger radii (and lower heights) in
the antiparallel array. This behavior can be understood by analyzing the
interdot dipolar coupling between two dots, $\tilde E(s)$. The absolute
value of the interaction energy increases with the core size, so $\tilde{E}%
^{u}(s)>\tilde{E}(s)$, as the uniform out-of-plane ($u$) configuration can
be seen as a vortex-core with an infinite core radius, and the magnetostatic
coupling for uniform out-of-plane magnetization is stronger than the
coupling of two vortex-cores. Therefore, the transition line in a parallel
array shifts to the left, because the interactions in an array of
vortex-core dots are less than the interactions in a uniformly magnetized P
array. On the other hand, in the AP arrays, the interactions are negative
and the uniformly magnetized AP configuration is favorable, so the complete
ferromagnetic ordering is reached for a smaller value of $h$.

\section{Conclusions}

We have examined the influence of dipolar interactions in the
effective core radius of dots in a vortex configuration, placed in
an $N\times N$ square array. Dipolar coupling among dots occurs
via core interaction, so we consider two types of relative
alignment between the cores: parallel, and antiparallel nearest
neighbors. Whenever a certain array configuration lowers the
interaction energy, the core region expands, as occurs in AP
interactions. The opposite behavior occurs for configurations
increasing the interaction energy, as the P case. These two
orderings represent a demagnetized configuration, the antiparallel
case, and a saturated one, corresponding to the parallel case.
Effects of the interaction between the dots in the core size have
been investigated by varying the number of dots in the array, the
distance between them, and their heights. In all cases, an
increase of the dipolar interaction energy leads to a decrease of
the core radius. When we increase the height of the dots, in the P
configuration a transition from vortex to a full ferromagnetic
state is hindered, while in the AP configuration the interaction
favors the transition to a full AP ferromagnetic state.

\begin{acknowledgments}
This work has been partially supported by Millennium Science
Nucleus "Basic and Applied Magnetism" P06-022F of Chile. MECESUP
USA0108 project and the program \textquotedblleft Bicentenario de
Ciencia y Tecnolog\'{\i}a (PBCT)\textquotedblright\ under the
project PSD-031 are also acknowledged. In Brazil the authors
acknowledge CNPq, PIBIC/UFRJ, FAPERJ, CAPES, PROSUL Program, and
Instituto de Nanotecnologia/MCT.
\end{acknowledgments}

\appendix

\section*{Appendix}

\setcounter{section}{1}.

The dipolar interaction energy between two identical dots $i$ and $j$ in the
vortex core configuration is given by%
\begin{equation*}
E_{dip}^{\pm }=\mu _{0}\int_{V_{i}}\vec{M}_{i}(\vec{r})\cdot \nabla U_{j}(%
\vec{r})\,dV^{\prime }
\end{equation*}%
\noindent where $\vec{M}_{i}$\ is the magnetization of the $i$-th\ dot and $%
U_{j}$\ is the magnetostatic potential of the $j$-th dot. For the vortex
core model defined by Eq.~(\ref{mv})we have $\nabla \cdot \vec{M}=0$, so the
magnetostatic potential reduces to\cite{LEA+05}%
\begin{multline}
U_{j}(\vec{r})=\frac{1}{4\pi }\sum_{p=-\infty }^{\infty
}\int\limits_{0}^{2\pi }e^{ip(\phi -\phi ^{\prime })}d\phi ^{\prime } \\
\int\limits_{0}^{R}M_{z}(\rho ^{\prime })J_{p}(k\rho ^{\prime })\rho
^{\prime }d\rho ^{\prime }\int\limits_{0}^{\infty }J_{p}(k\rho
_{j})[e^{-k(H-z)}-e^{-kz}]dk\,.
\end{multline}%
\noindent In this expression we have used the expansion\cite{Jackson75}%
\begin{equation}
\frac{1}{\left\vert \vec{r}-\vec{r}^{\prime }\right\vert }=\sum_{p=-\infty
}^{\infty }e^{ip(\phi -\phi ^{\prime })}\int_{0}^{\infty }J_{p}(k\rho
)J_{p}(k\rho ^{\prime })e^{-k(z_{>}-z_{<})}dk\;.
\end{equation}%
Here $J_{p}(z)$ are Bessel functions of first kind. The angular integration
gives us $\int_{0}^{2\pi }e^{ip(\phi -\phi ^{\prime })}d\phi ^{\prime }=2\pi
e^{ip\phi }\delta _{p,0}$, leading to%
\begin{multline}
U_{j}\left( \rho _{j},z\right) =\frac{1}{2}\int_{0}^{R}\rho ^{\prime
}M_{z}\left( \rho ^{\prime }\right) J_{0}(k\rho ^{\prime })d\rho ^{\prime }
\label{PF9} \\
\int_{0}^{\infty }J_{0}(k\rho _{j})\left[ e^{-k(H-z)}-e^{-kz}\right] dk.
\end{multline}%
\noindent Since the potential has no dependence on $\phi $, the expression
for the energy reduces to%
\begin{equation}
E^{\pm }=\mu _{0}\int\limits_{0}^{H}\int\limits_{0}^{2\pi
}\int\limits_{0}^{R}M_{z}(\rho )\frac{\partial U_{j}(\rho _{j},z)}{\partial z%
}\rho d\rho d\phi \,dz\,,  \label{EF}
\end{equation}%
\noindent and using Eq.~(\ref{PF9}) it is straightforward to obtain%
\begin{multline}
E^{\pm }=\mu _{0}\int\limits_{0}^{2\pi }d\phi \int\limits_{0}^{\infty
}dk\int\limits_{0}^{R}J_{0}(k\rho _{j})M_{z}(\rho )\rho d\rho   \label{int9}
\\
\int\limits_{0}^{R}J_{0}(k\rho ^{\prime })M_{z}(\rho ^{\prime
})(1-e^{-kH})\rho ^{\prime }d\rho ^{\prime }\,.
\end{multline}%
We need to evaluate the potential due to dot $j$ on dot $i$ a distance $S$
apart. Then we have to relate the radial coordinates of both dots through
the relation
\begin{equation*}
\rho _{j}=\sqrt{\rho ^{2}+S^{2}-2\rho S\cos (\phi +\beta )}\,,
\end{equation*}%
with $\beta $ an arbitrary angle. Using the following identity\cite%
{Jackson75}%
\begin{eqnarray*}
J_{0}(k\rho _{j}) &=&J_{0}\left( k\sqrt{\rho ^{2}+S^{2}-2\rho S\cos (\phi
+\beta )}\right)  \\
&=&\sum_{p=-\infty }^{\infty }e^{ip(\phi +\beta )}J_{p}(k\rho )J_{p}(kS)\,
\end{eqnarray*}%
in the expression for the energy (Eq. \ref{int9}),\ after the angular
integration, we obtain
\begin{equation}
\tilde{E}^{\pm }[S]=\pm \frac{2\pi }{L_{\mbox{\it\scriptsize x}}^{3}}%
\int\limits_{0}^{\infty }dk(1-e^{-kH})J_{0}(kS)\left[ \int%
\limits_{0}^{R}J_{0}(k\rho )m_{z}(\rho )\rho d\rho \right] ^{2}\,.
\end{equation}


\begin{thebibliography}{99}
\bibitem{V01} M. V\'{a}zquez, Physica B \textbf{299}, 302-313 (2001).

\bibitem{NHW+02} K. Nielsch, R. Hertel, R. B. Wehrspohn, J. Barthel, J.
Kirschner, U. G\"{o}sele, S. F. Fischer, and H. Kronm\"{u}ller, IEEE Trans.
Magn. \textbf{38}, 2571 (2002).

\bibitem{CKA+99} R. P. Cowburn, D. K. Koltsov, A. O. Adeyeye, M. E. Welland, and
D. M. Tricker, Phys. Rev. Lett. \textbf{83}, 1042 (1999).

\bibitem{RHS+02} C. A. Ross, M. Hwang, M. Shima, J. Y. Cheng, M. Farhoud, T. A. Savas,
Henry I. Smith, W. Schwarzacher, F. M. Ross, M. Redjdal, and F. B.
Humphrey, Phys. Rev. B \textbf{65}, 144417 (2002).

\bibitem{CRE+04} F. J. Casta\~{n}o, C. A. Ross, A. Eilez, W. Jung, and C. Frandsen, Phys. Rev. B \textbf{69}, 144421 (2004).

\bibitem{RKL+01} J. Rothman, M. Kl\"{a}ui, L. Lopez-Diaz, C. A. F. Vaz, A. Bleloch,
J. A. C. Bland, Z. Cui, and R. Speaks, Phys. Rev. Lett.
\textbf{86}, 1098 (2001).

\bibitem{NCR+05} K. Nielsch, F. J. Casta\~{n}o, C. A. Ross, and R. Krishnan, J. Appl. Phys. \textbf{98}, 034318 (2005).

\bibitem{NCM+05} Kornelius Nielsch, Fernando J. Casta\~{n}o, Sven
Matthias, Woo Lee, and Caroline A. Ross, Adv. Eng. Mat.
\textbf{7}, 217-221 (2005).

\bibitem{Chou97} S. Y. Chou, Proc. IEEE \textbf{85}, 652 (1997); G. Prinz, Science \textbf{282}, 1660 (1998).

\bibitem{Daughton99} J. M. Daughton, A. V. Pohm, R. T. Fayfield, and C. H. Smith, J. Phys. D \textbf{32}, R169 (1999).

\bibitem{ZZP00} J. G. Zhu, Y. Zheng, and Gary A. Prinz, J. App. Phys. \textbf{87}, 6668 (2000).

\bibitem{Parkin04} Stuart S. P. Parkin, Christian Kaiser, Alex Panchula, Philip M. Rice, Brian Hughes, Mahesh Samant, and See-Hun Yang, Nat.
Mater. \textbf{3}, 862 (2004).

\bibitem{AAR+02} J. d'Albuquerque e Castro, D. Altbir, J. C. Retamal, and P. Vargas, Phys. Rev. Lett. \textbf{88}, 237202 (2002).

\bibitem{LEA+05} P. Landeros, J. Escrig, D. Altbir, D. Laroze, J. d'Albuquerque e
Castro, and P. Vargas, Phys. Rev. B \textbf{71}, 094435 (2005).

\bibitem{PH04} F. Porrati, and M. Huth, Appl. Phys. Lett. \textbf{%
85}, 3157 (2004).

\bibitem{ELA+06} J. Escrig, P. Landeros, D. Altbir, M. Bahiana, and J. d'Albuquerque
e Castro, Appl. Phys. Lett. \textbf{89}, 132501 (2006).

\bibitem{Metlov06} Konstantine L. Metlov, Phys. Rev. Lett. \textbf{97}, 127205 (2006).

\bibitem{LEL+07} D. Laroze, J. Escrig, P. Landeros, D. Altbir, M. V\'{a}zquez,
and P. Vargas, Nanotechnology \textbf{18}, 415708 (2007).

\bibitem{EAJ+07} J. Escrig, D. Altbir, M. Jaafar, D. Navas, A. Asenjo, and M. V\'{a}%
zquez, Phys. Rev. B \textbf{75}, 184429 (2007).

\bibitem{PH05} F. Porrati, and M. Huth, J. Magn. Magn. Mater. \textbf{290}, 145-148 (2005).

\bibitem{SOH+00} T. Shinjo, T. Okuno, R. Hassdrof, K. Shigeto, and T. Ono, Science \textbf{289}, 930 (2000).

\bibitem{WWB+02} A. Wachowiak, J. Wiebe, M. Bode, O. Pietzsch, M. Morgenstern, and
R. Wiesendanger, Science \textbf{298}, 577 (2002).

\bibitem{Aharoni96} A. Aharoni, Introduction to the Theory of
Ferromagnetism (Oxford: Clarendon, 1996).

\bibitem{RSpc} I. V. Roshchin, and I. K. Schuller, private comm.

\bibitem{FT65} E. Feldtkeller, and H. Thomas, Phys. Kondens. Mater. \textbf{4}, 8 (1965); P. O. Jubert, and R. Allenspach, Phys. Rev. B \textbf{70}, 144402 (2004).

\bibitem{UP93} N. A. Usov, and S. E. Peschany, J. Magn. Magn. Mater. \textbf{118}, L290 (1993).

\bibitem{Aharoni90} A. Aharoni, J. Appl. Phys. \textbf{68},
2892-2900 (1990).

\bibitem{HKK03} R. H\"{o}llinger, A. Killinger, and U. Krey, J.
Magn. Magn. Mater. \textbf{261}, 178-189 (2003).

\bibitem{EM04} O. Espinosa, and V. H. Moll, Integral Transforms and
Special Functions \textbf{15}, 101-115 (2004).

\bibitem{TBZ+04} S. Tandon, M. Beleggia, Y. Zhu, and M. De Graef, J.
Magn. Magn. Mater. \textbf{271}, 21 (2004).

\bibitem{BTZ+04} M. Beleggia, S. Tandon, Y. Zhu, and M. De Graef, J.
Magn. Magn. Mater. \textbf{278}, 270 (2004).

\bibitem{Jackson75} J. D. Jackson, Classical Electrodynamics, 2nd
Edition (John Wiley \& Sons, 1975).
\end{thebibliography}
\end{document}